\documentclass[fleqn,10pt]{wlscirep}

\usepackage{bm}
\usepackage{fixmath}
\usepackage{amsmath,amssymb}

\title{Tunable magnetoplasmonics in lattices of Ni/SiO$_2$/Au dimers}

\author[1]{Sara Pourjamal}
\author[1,2]{Mikko Kataja}
\author[3]{Nicol\`{o} Maccaferri}
\author[4]{Paolo Vavassori}
\author[1,*]{Sebastiaan van Dijken}
\affil[1]{NanoSpin, Department of Applied Physics, Aalto University School of Science, P.O. Box 15100, FI-00076 Aalto, Finland}
\affil[2]{Institut de Ci\`{e}ncia de Materials de Barcelona (ICMAB-CSIC), Campus de la UAB, Bellaterra, Catalonia, Spain}
\affil[3]{Istituto Italiano di Tecnologia, Via Morego 30, 16163 Genova, Italy}
\affil[4]{CIC nanoGUNE, E-20018 Donostia-San Sebastian,	Spain; and Ikerbasque, Basque Foundation for Science, E-48013 Bilbao, Spain}
\affil[*]{sebastiaan.van.dijken@aalto.fi}

\begin{abstract}
We present a systematic study on the optical and magneto-optical properties of Ni/SiO$_2$/Au dimer lattices. By considering the excitation of orthogonal dipoles in the Ni and Au nanodisks, we analytically demonstrate that the magnetoplasmonic response of dimer lattices is governed by a complex interplay of near- and far-field interactions. Near-field coupling between dipoles in Ni and low-loss Au enhances the polarizabilty of single dimers compared to that of isolated Ni nanodisks. Far-field diffractive coupling in periodic lattices of these two particle types enlarges the difference in effective polarizability further. This effect is explained by an inverse relationship between the damping of collective surface lattice resonances and the imaginary polarizability of individual scatterers. Optical reflectance measurements, magneto-optical Kerr effect spectra, and finite-difference time-domain simulations confirm the analytical results. Hybrid dimer arrays supporting intense plasmon excitations are a promising candidate for active magnetoplasmonic devices.
\end{abstract}
\begin{document}

\flushbottom
\maketitle
\thispagestyle{empty}

\section*{Introduction}
Noble-metal nanoparticles are widely used in plasmonics because their high electrical conductivity supports the excitation of low-loss localized surface plasmon resonances (LSPRs)\cite{MAI-07}. The ensuing optical response of metal nanoparticles can be tuned by variation of their size, shape, or arrangement\cite{BRY-08,SCH-10}. Strong enhancements of the optical field at the surface of metal nanoparticles and in their immediate vicinity are exploited, for instance, in biological and chemical sensors\cite{MAY-08,VER-11}, photovoltaics\cite{ERW-16}, and optoelectronics\cite{PAC-07}. Nanoparticles made of ferromagnetic metals also support the excitation of LSPRs\cite{CHE-11,BON-11,MAC-13,MAC-15}. Since plasmon resonances and magneto-optical activity are strongly linked in ferromagnetic nanoparticles, their magneto-optical spectra can be tailored by employing design rules known from plasmonics. Conversely, nanoscale ferromagnets enable active control of light via magnetization reversal in an external field. Both effects are relevant for technology and are studied in the field of magnetoplasmonics\cite{ARM-13,BOS-16,FLO-18}. 

Large ohmic losses in ferromagnetic metals lead to significant damping of plasmon resonances. To overcome this limitation, hybrid structures comprising ferromagnetic and noble metals have been explored as magnetoplasmonic systems. Examples include, Au/Co/Au trilayers\cite{TEM-10}, nanosandwiches\cite{GON-08}, and nanorods \cite{ARME-16}, core-shell Co/Ag or Co/Au nanoparticles\cite{WAN-11,SON-12} and nanowires\cite{TOA-14}, and Au/Ni nanoring resonators\cite{ATM-18}. Contacting subwavelength ferromagnetic elements and noble metals results in materials that can be considered as optical alloys. Various non-contacting realizations have also been investigated. Dimers of two metal nanodisks that are separated by a dielectric layer are particularly attractive as they allow for a strong redistribution of the optical near-field\cite{NOR-04}. In vertical dimers containing noble and ferromagnetic metals, this effect has been exploited to enlarge the magneto-optical response via an increase of the optical field in the ferromagnetic layer\cite{BAN-12} or induction of magneto-optical activity on the low-loss noble metal\cite{ARM-13-2,SOU-14}. 

Another way to circumvent large ohmic losses in ferromagnetic nanoparticles involves the excitation of collective plasmon modes. In periodic arrays of metal nanoparticles, constructive interference of the optical fields from individual scatterers produces narrow and intense surface lattice resonances (SLRs)\cite{KRA-08,AUG-08,HUM-14}. Low-loss SLRs in arrays of noble metal nanostructures are used in several contexts, including sensing\cite{OFF-11,SHE-13,LOD-13}, lasing\cite{ZHO-13,HAK-17}, and metamaterials\cite{ZHA-12,POR-13}. In ferromagnetic nanoparticle arrays, SLRs enhance the magneto-optical activity and provide versatility in the design of magneto-optical spectra via tailoring of lattice symmetry or particle shape\cite{KAT-15,MAC-16}. Checkerboard patterns of pure Ni and Au nanodisks have also been studied\cite{KAT-16}. In this hybrid approach, far-field diffractive coupling between the different particles enhances the magneto-optical response via the excitation of low-loss SLRs and induction of magneto-optical activity on the Au nanodisks.  

\begin{figure}[t]
	\centering
	\includegraphics[width=0.5\linewidth]{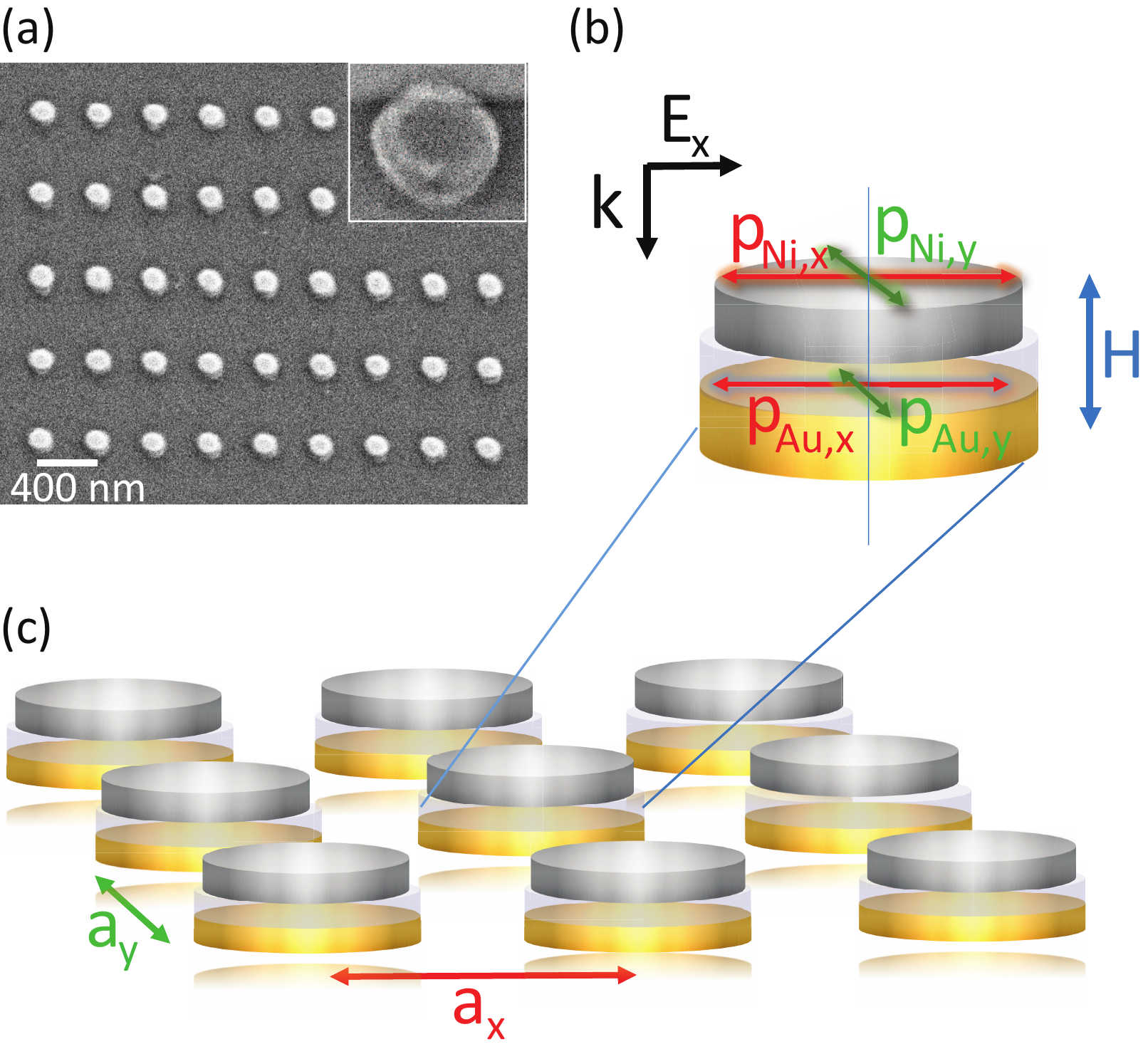}
	\caption{(a) Scanning electron microscopy (SEM) image and (b,c) schematic of a Ni/SiO$_2$/Au dimer lattice. The dimers are patterned onto a glass substrate with Au nanodisks at the bottom and Ni nanodisks at the top. The two metal disks are separated by 15 nm SiO$_2$. Optical and magneto-optical measurements are performed with linearly polarized light at normal incidence ($E$-field along $x$). A perpendicular magnetic field ($H$) saturates the magnetization of the Ni nanodisks. We study dimer arrays with different lattice constants ($a_\mathrm{x}$, $a_\mathrm{y}$) and compare the results to those measured on arrays with Au and Ni nanodisks.}
	\label{fig:1}
\end{figure}

Here, we report on tunable magnetoplasmonics in lattices of Ni/SiO$_2$/Au dimers (Fig. \ref{fig:1}). Our structures combine two aforementioned approaches, namely, the integration of noble and ferromagnetic metals in vertical dimers\cite{BAN-12,ARM-13-2,SOU-14} and ordering of magneto-optically active elements in periodic arrays\cite{KAT-15,MAC-16}. Because the noble metal and ferromagnetic constituents of our lattices interact via optical near-fields within dimers and far-fields between dimers, the hybrid arrays provide a rich playground for the design of optical and magneto-optical effects. First, we present an analytical model to evaluate the effect of dimer polarizability and lattice periodicity on the magnetoplasmonic properties of our system. Next, we compare model calculations and experiments on dimer arrays with different lattice constants. As reference, we discuss experiments on arrays with Au and Ni nanodisks.             

\section*{Modeling}
We start our analysis by calculating the optical and magneto-optical response of an individual plasmonic nanoparticle based on the modified long wavelength approximation (MLWA)\cite{MOR-09}. The absorption and emission properties of a metal nanoparticle are described by its volume polarizability $\alpha_\mathrm{e}'$, which relates the induced polarization $\boldsymbol{P}$ to the incident electric field $\boldsymbol{E}_\mathrm{i}$. If the particle is small compared to the wavelength of incident light, the electric field inside the particle $\boldsymbol{E}_1$ is approximately uniform. Following classical electrodynamics, the electric field inside the nanoparticle is given by $\boldsymbol{E}_1=\boldsymbol{E}_\mathrm{i}+\boldsymbol{E}_\mathrm{d}$, where $\boldsymbol{E}_\mathrm{d}$ is the depolarization field. $\boldsymbol{E}_\mathrm{d}$ can be calculated by assigning a dipole moment $d\boldsymbol{p}=\boldsymbol{P}dV$ to each volume element $dV$ of the nanoparticle and calculating the retarded depolarization field $d\boldsymbol{E}_\mathrm{d}$ generated by $d\boldsymbol{p}$ in the nanoparticle center\cite{MAC-13-2}. This gives 

\begin{equation}
{\boldsymbol{E}_\mathrm{d} = \int d\boldsymbol{E}_\mathrm{d} = -L_\mathrm{d}\boldsymbol{P}}.
\label{depolarization-field}
\end{equation}

\noindent Here, $L_\mathrm{d}$ is the depolarization factor describing interactions between polarizable volume elements inside the particle\cite{BAR-82}. The nanodisks that we consider in our study can be approximated as ellipsoids\cite{MOR-09,BOH-83}. For ellipsoidal particles, Eq. \ref{depolarization-field} can be solved analytically. This gives

\begin{equation}
{L_\mathrm{d} = L-\frac{ik^3V}{6\pi}I-\frac{k^2V}{4\pi}D}.
\label{depolarization-tensor}
\end{equation}

\noindent The three terms in Eq. \ref{depolarization-tensor} include static ($L$) and dynamic ($D$) depolarization factors that account for the particle shape and a radiative reaction correction ($ik^3V/6\pi$)\cite{MAC-13-2}. To calculate $L$ and $D$, we use the integrals given in Refs. \citenum{MOR-09,MAC-13-2}. The net dipole moment of an ellipsoidal particle ($d\boldsymbol{p}=\boldsymbol{P}dV$) can be written as	

\begin{equation}
{\boldsymbol{p} = (\epsilon_\mathrm{d}-\epsilon_\mathrm{m})\boldsymbol{E}_{1}V = (\epsilon_\mathrm{d}-\epsilon_\mathrm{m})(\boldsymbol{E}_\mathrm{i}+\boldsymbol{E}_\mathrm{d})V = \alpha_\mathrm{e}\boldsymbol{E}_\mathrm{i}}, 
\label{dipole-moment}
\end{equation}

\noindent where $\epsilon_\mathrm{d}$ and $\epsilon_\mathrm{m}$ are the permittivity of the particle and surrounding medium, respectively, $\alpha_\mathrm{e}$ is the particle polarizability ($\alpha_\mathrm{e}=\alpha_\mathrm{e}'V$), and $V$ is the particle volume. Combining Eqs. \ref{depolarization-field} and \ref{dipole-moment} gives

\begin{equation}
{\alpha_\mathrm{e} = \frac{(\epsilon_\mathrm{d}-\epsilon_\mathrm{m})}{I+L_\mathrm{d}\epsilon_\mathrm{m}^{-1}(\epsilon_\mathrm{d}-\epsilon_\mathrm{m})}V}. 
\label{polarizability}
\end{equation}

\noindent The permittivity of a particle changes in the presence of a large external magnetic field or spontaneous magnetization. In our experiments, we use perpendicular magnetic fields of $\pm400$ mT to saturate the magnetization of Ni nanodisks along the $z$-axis. The permittivity tensor for this configuration contains two off-diagonal components\cite{ZVE-97}

\begin{equation}
\renewcommand*{\arraystretch}{1.2}
\epsilon_\mathrm{d} = 
\begin{pmatrix}
\epsilon_\mathrm{xx} & -iQm_\mathrm{z} & 0 \\
iQm_\mathrm{z} & \epsilon_\mathrm{yy} & 0 \\
0 & 0 & \epsilon_\mathrm{yy}.
\end{pmatrix}\,,
\label{permittivity-tensor}
\end{equation}

\noindent where $m_\mathrm{z}$ is the perpendicular magnetization and $Q$ is the Voigt magneto-optical constant. We use tabulated data from Ref. \citenum{VIS-93} to calculate the permittivity of Ni. Because the field-induced diamagnetic moment of Au is small ($m_\mathrm{z}\approx0$) compared to the magnetization of Ni, we set the off-diagonal terms of $\epsilon_\mathrm{d}$ to zero for this material. We use optical constants from Ref. \citenum{JOH-72} to calculate the permittivity of Au. 

Following Eq. \ref{polarizability}, non-zero off-diagonal components in $\epsilon_\mathrm{d}$ lead to off-diagonal terms in the polarizability tensor. Macroscopically, this produces a rotation and ellipticity in the polarization of reflected (magneto-optical Kerr effect) or transmitted (Faraday effect) light. For nanoparticles, the microscopic origin of magneto-optical activity can be understood by considering the excitation of two orthogonal LSPRs. One of the LSPRs, which can be described as electric dipole $\boldsymbol{p}$, is driven by the incident electric field $\boldsymbol{E}_\mathrm{i}$. For linearly polarized light at normal incidence, the induced dipole is oriented in-plane along $\boldsymbol{E}_\mathrm{i}$. If the nanoparticle exhibits perpendicular magnetization ($m_\mathrm{z}$), a second electric dipole is induced orthogonal to $\boldsymbol{E}_\mathrm{i}$ and $m_\mathrm{z}$ by spin-orbit coupling. The amplitude and phase relations of the two excited dipoles determine the rotation and ellipticity of light polarization upon reflection or transmission\cite{MAC-13}. In our study, the incident electric field is oriented along the $x$-axis, the magnetization of Ni is saturated by a perpendicular magnetic field, and the spin-orbit induced dipole is oriented along $y$ (Fig. \ref{fig:1}(b)). Hereafter, we refer to the directly excited dipole ($p_\mathrm{x}$) as optical dipole. The dipole along the orthogonal direction ($p_\mathrm{y}$) is labeled as magneto-optical dipole. 

If dimers are formed from Au and Ni nanodisks, their optical near-fields couple. To describe this effect, we consider the electric field at each dipole position as the sum of the incident electric field and the scattered field from the dipole in the other disk. This results in two coupled equations

\begin{equation}
	\begin{aligned}
	\boldsymbol{p}_\mathrm{Ni} = \alpha_\mathrm{Ni}(\epsilon_0\boldsymbol{E}_\mathrm{i1}+\boldsymbol{G}\boldsymbol{p}_\mathrm{Au}), \\
	\boldsymbol{p}_\mathrm{Au} = \alpha_\mathrm{Au}(\epsilon_0\boldsymbol{E}_\mathrm{i2}+\boldsymbol{G}\boldsymbol{p}_\mathrm{Ni}).
	\end{aligned}
\label{dipole-Ni-Au}
\end{equation}

\noindent Here, $\boldsymbol{E}_\mathrm{i1}$ and $\boldsymbol{E}_\mathrm{i2}$ define the incident electric field at the Ni and Au nanodisks (including a phase difference), and $\boldsymbol{G}$ is a dyadic Green’s function describing how the electric field that is produced by one dipole propagates to the other \cite{DAP-94}. $\boldsymbol{G}$ is given by

\begin{equation}
{\boldsymbol{G} = \frac{e^{ikR}}{4{\pi}\epsilon_{0}R^3}\Big(\big((kR)^2 + ikR - 1\big)I - \big((kR)^2 + 3ikR - 3\big)\frac{\boldsymbol{R}{\otimes}\boldsymbol{R}}{R^2}\Big)}, 
\label{G-factor}
\end{equation}

\noindent where $\boldsymbol{R}$ is a vector connecting the dipoles in the two disks, $R$ is its amplitude, and $k=2n\pi/\lambda$, with $n$ the refractive index of the spacer layer and surrounding medium. Since electric dipoles are excited in the dimer plane, they mostly couple along the $z$-axis. Consequently, $\boldsymbol{R}{\otimes}\boldsymbol{R}$ in Eq. \ref{G-factor} is approximately zero. The optical and magneto-optical spectra of dimers are defined by dipole excitations along $x$ and $y$. Considering near-field coupling between the Ni and Au nanodisks (Eq. \ref{dipole-Ni-Au}), the effective dipole moment along these axes can be written as

\begin{equation}
\renewcommand*{\arraystretch}{1.2}
\begin{pmatrix}
p_\mathrm{x} \\
p_\mathrm{y} 
\end{pmatrix} =
\begin{pmatrix}
p_\mathrm{Ni,x} + p_\mathrm{Au,x} \\
p_\mathrm{Ni,y} + p_\mathrm{Au,y}
\end{pmatrix} = 
\begin{pmatrix}
\alpha_\mathrm{xx} & -\alpha_\mathrm{xy} \\
\alpha_\mathrm{xy} & \alpha_\mathrm{yy}
\end{pmatrix}
\begin{pmatrix}
E_\mathrm{x} \\
0
\end{pmatrix},
\label{dimer-dipole}
\end{equation}

\noindent where $\alpha_\mathrm{xx}$, $\alpha_\mathrm{yy}$, and $\alpha_\mathrm{xy}$ are the diagonal and off-diagonal components of the polarizability tensor ($\alpha$) of a single Ni/SiO$_2$/Au dimer. We note that while off-diagonal components are absent in the polarizability matrix of Au, a magneto-optical dipole is induced on the Au nanodisk ($p_\mathrm{Au,y}$) because of near-field coupling to $p_\mathrm{Ni,y}$ (Eq. \ref{dipole-Ni-Au}). The low-loss Au nanodisk thus contributes to the magneto-optical activity of the dimer\cite{ARM-13-2,SOU-14}.    

If dimers are ordered into a periodic array, the electric field at each lattice position is a superposition of the incident radiation and dipolar fields from other dimers. The optical and magneto-optical response of a periodic dimer array thus depend on the polarizability of single dimers ($\alpha$) and their two-dimensional arrangement. To take far-field coupling between dimers into account, we define an effective lattice polarizability\cite{GAR-07,ZOU-04}

\begin{equation}
\alpha_\mathrm{eff} = \frac{1}{1/\alpha - S}, 
\label{effective-polarizability}
\end{equation}

\noindent where $S$ is the lattice factor. For an infinite array, this parameter is given by\cite{DEJ-14,EVL-12}

\begin{equation}
S = \sum_j e^{ikr_j}\Big(\frac{(1-ikr_j)(3\cos^2(\theta_j) - 1)}{{r_j}^3} + \frac{k^2\sin^2(\theta_j)}{r_j}\Big),
\label{lattice-factor}
\end{equation}

\noindent where $r_j$ is the distance between dimers and $\theta_j$ is the angle between the effective dipole moment and the vector connecting the dimers. For a two-dimensional lattice under normal incidence radiation, we can thus write

\begin{equation}
\renewcommand*{\arraystretch}{1.2}
\alpha_\mathrm{eff} = \Bigg(
\begin{pmatrix}
\alpha_\mathrm{xx} & -\alpha_\mathrm{xy} \\
\alpha_\mathrm{xy} & \alpha_\mathrm{yy} 
\end{pmatrix}^{-1} -
\begin{pmatrix}
S_\mathrm{x} & 0 \\
0 & S_\mathrm{y}
\end{pmatrix}\Bigg)^{-1},
\label{effective-polarizability2}
\end{equation}

\noindent where $S_\mathrm{x}$ and $S_\mathrm{y}$ are the lattice factors for radiation along $x$ and $y$. Since $\alpha_\mathrm{xx,yy}>>\alpha_\mathrm{xy}$, the diagonal components of the effective lattice  polarizability only depend on the diagonal terms of $\alpha$ and $S$. The off-diagonal components of $\alpha_\mathrm{eff}$ contain more intricate parameter relations. By carrying out matrix operations (see Supplementary Note 1), we find

\begin{equation}
\begin{aligned}
\alpha_\mathrm{eff,xx} = \frac{1}{1/\alpha_\mathrm{xx} - S_\mathrm{x}},  \\
\alpha_\mathrm{eff,yy} = \frac{1}{1/\alpha_\mathrm{yy} - S_\mathrm{y}}, 
\end{aligned}
\label{effective-polarizability-xx}
\end{equation}

\noindent and

\begin{equation}
\alpha_\mathrm{eff,xy} = \frac{\alpha_\mathrm{xy}}{\alpha_\mathrm{xx}\alpha_\mathrm{yy}(1/\alpha_\mathrm{yy} - S_\mathrm{x})(1/\alpha_\mathrm{xx} - S_\mathrm{y})}. 
\label{effective-polarizability-xy}
\end{equation}

\noindent The effective dipole moments of the dimer lattice are thus given by

\begin{equation}
\renewcommand*{\arraystretch}{1.2}
\begin{pmatrix}
p_\mathrm{eff,x} \\
p_\mathrm{eff,y} 
\end{pmatrix} =
\begin{pmatrix}
\alpha_\mathrm{eff,xx} & -\alpha_\mathrm{eff,xy} \\
\alpha_\mathrm{eff,xy} & \alpha_\mathrm{eff,yy}
\end{pmatrix}
\begin{pmatrix}
E_\mathrm{x} \\
0
\end{pmatrix},
\label{lattice-dipole}
\end{equation} 

\noindent Equation \ref{effective-polarizability-xy} reveals a complex relationship between the polarizability of the dimers and their periodic arrangement. Because magneto-optical dipoles ($p_\mathrm{y}$) are excited orthogonal to the optical dipoles ($p_\mathrm{x}$), the polarizability and lattice factor along the $y$-axis also affect $\alpha_\mathrm{eff,xy}$\cite{MAC-16}. 

For linearly polarized light at normal incidence, the optical reflectance and magneto-optical activity are linked simply to the effective lattice polarizability. In this geometry, the reflectance of a periodic plasmonic array is proportional to the scattering cross section\cite{BOH-83}

\begin{equation}
R \propto \sigma_\mathrm{sca} = \frac{k^4}{6\pi}|\alpha_\mathrm{eff,xx}|^2,
\label{reflectivity}
\end{equation}

\noindent and thus

\begin{equation}
R \propto |p_\mathrm{eff,x}|^2.
\label{reflectivity-dipole}
\end{equation}

\noindent The magneto-optical Kerr angle $\Phi$ of a dimer lattice is defined as the amplitude ratio of the magneto-optical ($p_\mathrm{eff,y}$) and optical ($p_\mathrm{eff,x}$) dipoles

\begin{equation}
\Phi = \Bigg|\frac{p_\mathrm{eff,y}}{p_\mathrm{eff,x}}\Bigg| = \Bigg|\frac{\alpha_\mathrm{eff,xy}}{\alpha_\mathrm{eff,xx}}\Bigg|. 
\label{Kerr-angle}
\end{equation}

\noindent Following Eqs. \ref{reflectivity-dipole} and \ref{Kerr-angle}, it is possible to extract a quantity that is proportional to $|p_\mathrm{eff,y}|$ by multiplying the Kerr angle $\Phi$ by the square root of the optical reflectance $R$

\begin{equation}
|p_\mathrm{eff,y}|\propto\Phi\sqrt{R}. 
\label{MO-dipole}
\end{equation}

\section*{Results and Discussion}
To experimentally explore near- and far-field coupling in dimer arrays, we fabricated periodic lattices of Ni/SiO$_2$/Au dimers on glass substrates using electron-beam lithography\cite{POU-18}. The lower Au and upper Ni nanodisks of the dimers have a diameter of $\sim$120 nm and $\sim$110 nm, respectively, and both disks are 15 nm thick. The two metals are separated by 15 nm SiO$_2$. The lattice constant along $x$ and $y$ are 400 nm, 450 nm, or 500 nm. For comparison, we also fabricated arrays of pure Au and Ni nanodisks. The Au nanodisks have the same size as in the dimers. Because the optical reflectance from pure Ni nanodisks is small, we decided to increase their diameter and thickness to $\sim$130 nm and 18 nm. In addition, we fabricated samples with randomly distributed dimers and nanodisks to characterize the optical and magneto-optical response without SLRs. All measurements were conducted with the nanoparticles immersed in index-matching oil ($n=1.52$). The creation of a homogeneous refractive-index environment enhances the efficiency of far-field coupling between scatterers and, thereby, the excitation of collective SLR modes. More experimental details are given in the Methods section.

\begin{figure}[t]
	\centering
	\includegraphics[width=\linewidth]{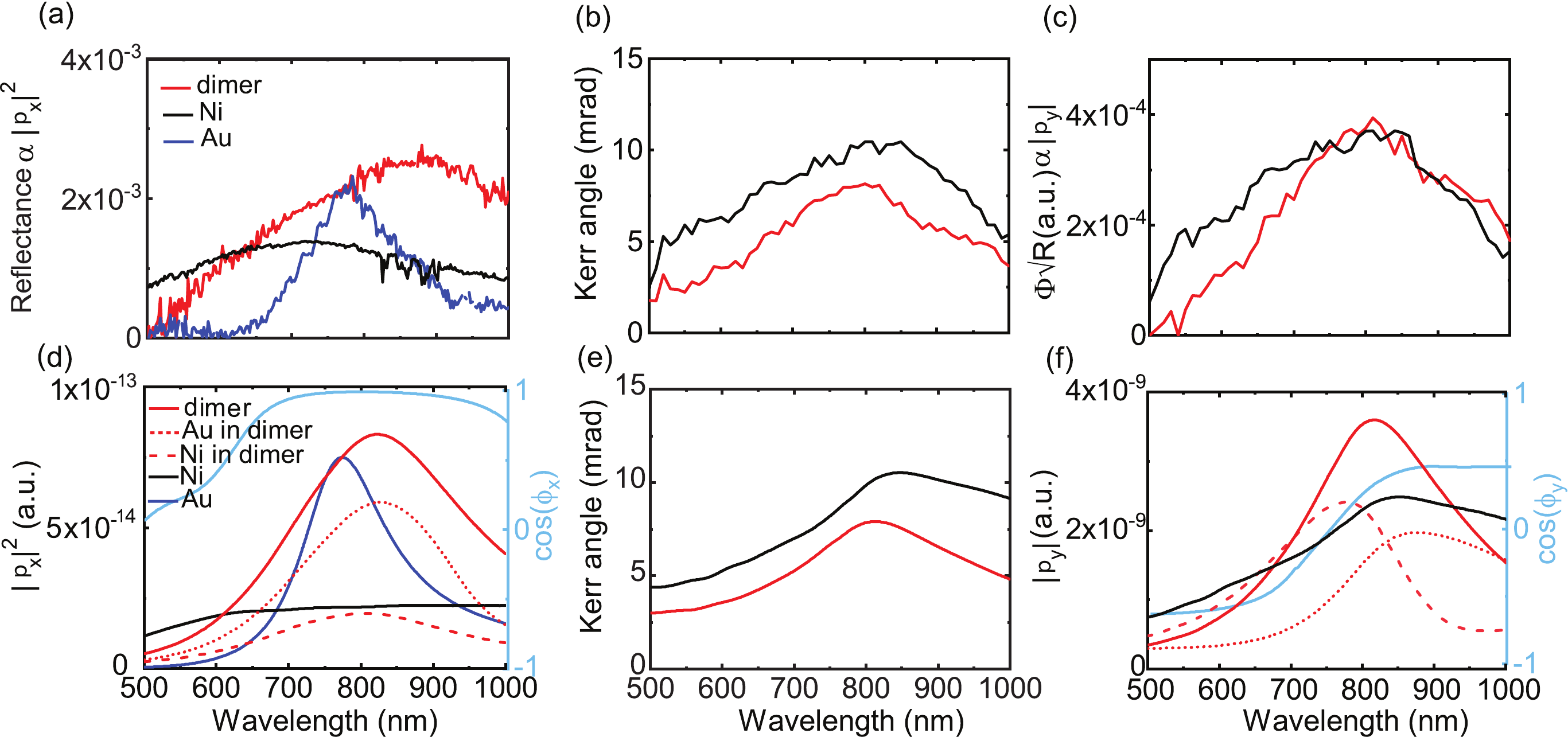}
	\caption{(a) Optical reflectance ($R$) of randomly distributed Ni/SiO$_2$/Au dimers, Ni nanodisks, and Au nanodisks. (b,c) Measured Kerr angle ($\Phi$) and extracted values of $\Phi\sqrt{R}$ for samples with random dimers and Ni nanodisks. The parameter in (c) is proportional to the magneto-optical dipole amplitude ($|p_\mathrm{y}|$). (d-f) Calculations of $|p_\mathrm{x}|^2$, the magneto-optical Kerr angle ($|p_{y}/p_\mathrm{x}|$), and $|p_{y}|$ for the same nanoparticles. In (d) and (f), the strengths of excited dipoles in the Au and Ni nanodisks of the dimer and their vector sum are plotted separately. These parameter are linked by Eq. \ref{dimer-dipole-moment}. Cosines of the phase difference between dipoles in Au and Ni are depicted in (d) and (f).}
	\label{fig:2}
\end{figure} 

\begin{figure}[t]
	\centering
	\includegraphics[width=\linewidth]{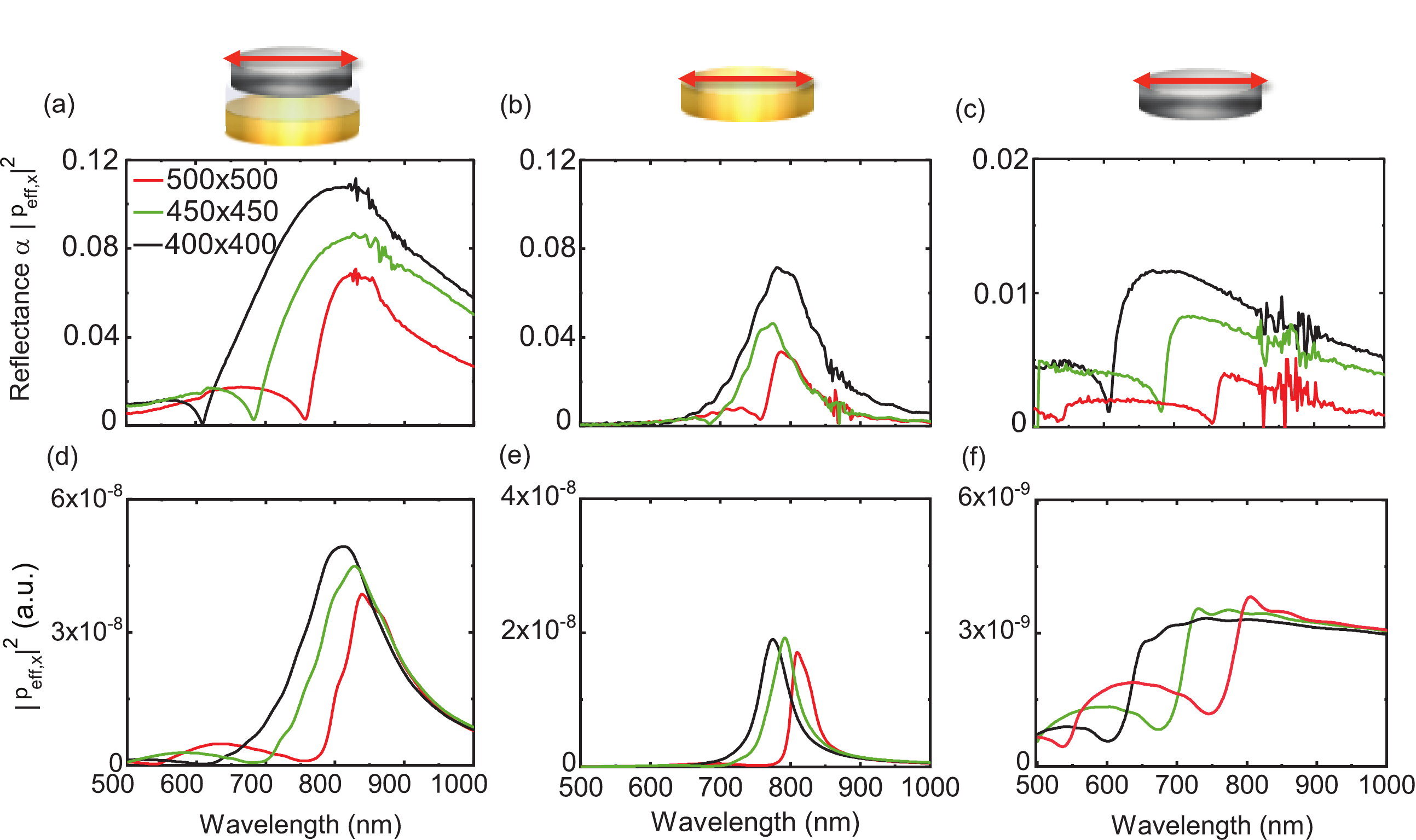}
	\caption{Optical reflectance ($R$) of square arrays of (a) Ni/SiO$_2$/Au dimers, (b) Au nanodisks, and (c) Ni nanodisks for three lattice constants. (d-f) Corresponding calculations of $|p_\mathrm{eff,x}|^2$ for the same lattices.}
	\label{fig:3}
\end{figure}

We first discuss the optical and magneto-optical response of randomly distributed dimers and nanodisks (Fig. \ref{fig:2}). A filling fraction of 5\% was chosen for these samples to approximately match those of periodic arrays (7\% for $a=400$ nm, 5\% for $a=500$ nm). Because of the low filling fraction, randomly distributed dimers and nanodisks can be considered as non-interacting and, consequently, their optical spectra represent the properties of individual nanoparticles. Figure \ref{fig:2}(a) compares reflectance spectra of randomly distributed Ni/SiO$_2$/Au dimers and Au and Ni nanodisks. Near-field coupling between the Au and Ni disks of dimers red-shifts the LSPR-induced reflectance maximum. The LSPR wavelength of a dimer is measured at $\sim$860 nm, while those of the Au and Ni nanosdisks are recorded at $\sim$790 nm and $\sim$720 nm, respectively. The LSPR linewidth of the dimer structure is also larger than that of the Au nanodisk because of dipolar coupling to a higher-loss excitation in Ni. Figure \ref{fig:2}(b) shows the magneto-optical Kerr angle of the dimer and Ni nanodisk. From data in Figs. \ref{fig:2}(a,b) we also extract $\Phi\sqrt{R}$, which is proportional to the magneto-optical dipole amplitude $|p_\mathrm{y}|$ (Eq. \ref{MO-dipole}). For the dimer structure (red line), $|p_\mathrm{y}|$ is the vector sum of a spin-orbit induced magneto-optical dipole in Ni ($p_\mathrm{Ni,y}$) and the dipole moment that it produces on Au ($p_\mathrm{Au,y}$). The values of $|p_\mathrm{y}|$ for the dimer and Ni nanodisk are similar at $\sim$800 nm, despite the latter containing $\sim$70\% more Ni. This result confirms that the Au nanodisk of a dimer contributes to the magneto-optical activity. We also note that $|p_\mathrm{y}|$ of the dimer structure decays more strongly below the resonance wavelength. This effect is caused by a weakening of the near-field coupling strength at shorter wavelengths, i.e., a decrease of $p_\mathrm{Au,y}$, as illustrated by calculations of the dyadic Green’s function describing dipolar coupling inside the dimer (Supplementary Note 2). 

To further delve into the details of near-field coupling in our magnetoplasmonic dimers, we present calculations of $|p_\mathrm{x}|^2$ and $|p_{y}|$ of single nanodisks and dimers in Figs. \ref{fig:2}(d,f). By plotting data in this format, the results can be compared directly to the experimental spectra of Figs. \ref{fig:2}(a,c). We also show the calculated magneto-optical Kerr angle ($|p_\mathrm{y}/p_\mathrm{x}|$) in Fig. \ref{fig:2}(e). In all cases, the wavelengths and lineshapes of plasmon resonances agree well. Main features such as a red-shift of the dimer LSPR are thus reproduced. In the calculations, we can separate how dipole moments in the Au and Ni nanodisks contribute to the optical and magneto-optical response of dimers. Taking the phase difference between excitations in Au and Ni along $x$ and $y$ ($\phi_\mathrm{x}$, $\phi_\mathrm{y}$) into account, the optical and magneto-optical dipoles of dimers are given by

\begin{equation}
\begin{aligned}
|p_\mathrm{x}|^2 = |p_\mathrm{Ni,x} + p_\mathrm{Au,x}|^2 = |p_\mathrm{Ni,x}|^2 + |p_\mathrm{Au,x}|^2 + 2|p_\mathrm{Ni,x}||p_\mathrm{Au,x}|\cos(\phi_\mathrm{x}),  \\
|p_\mathrm{y}|^2 = |p_\mathrm{Ni,y} + p_\mathrm{Au,y}|^2 = |p_\mathrm{Ni,y}|^2 + |p_\mathrm{Au,y}|^2 + 2|p_\mathrm{Ni,y}||p_\mathrm{Au,y}|\cos(\phi_\mathrm{y}).
\end{aligned}
\label{dimer-dipole-moment}
\end{equation}

\noindent Analyzing the results of Fig. \ref{fig:2}(f), we find that, in dimers, the maximum magneto-optical dipole strength in Au is about 75\% compared to that of Ni. The strong $p_\mathrm{Au,y}$ is explained by the large polarizability of Au, enabling $p_\mathrm{Ni,y}$ to effectively induce a magneto-optical dipole moment on Au. The calculations thus confirm the big impact of $p_\mathrm{Au,y}$ on the magneto-optical activity of single dimers.

We now consider far-field diffractive coupling in dimer lattices. Optical fields from individual scatterers in periodic arrays produce collective SLRs and narrow diffracted orders (DOs) in far-field measurements. The DO wavelengths are given by 

\begin{equation}
\sin\theta_k=\sin\theta_i+k\frac{\lambda}{na}, 
\label{DO}
\end{equation}

\noindent where $\theta_\mathrm{k}$ is the angle of the $k^{\mathrm{th}}$ diffracted order, $\theta_\mathrm{i}$ is the angle of incidence, $\lambda$ is the wavelength, $n$ is the refractive index of the embedding medium, and $a$ is the lattice constant. For normal incident light ($\theta_\mathrm{i}=0^\circ$), a Rayleigh anomaly associated with the passing of a DO is measured in reflectance or transmittance spectra when $k\lambda=na$. This corresponds to a transition from an evanescent to a propagating lattice mode if $\sin\theta_k=\pm1$ in Eq. \ref{DO}. For a two-dimensional lattice, the wavelengths of Rayleigh anomalies ($\lambda_\mathrm{DO}$) can be calculated using $\sqrt{(p^2+q^2)}\lambda_\mathrm{DO}=na$, where $p$ and $q$ indicate the order of diffraction along $x$ and $y$. Coupling of a DO with the broader LSPRs of individual nanoparticles produces a SLR with an asymmetric line-shape.

\begin{figure}[t]
	\centering
	\includegraphics[width=\linewidth]{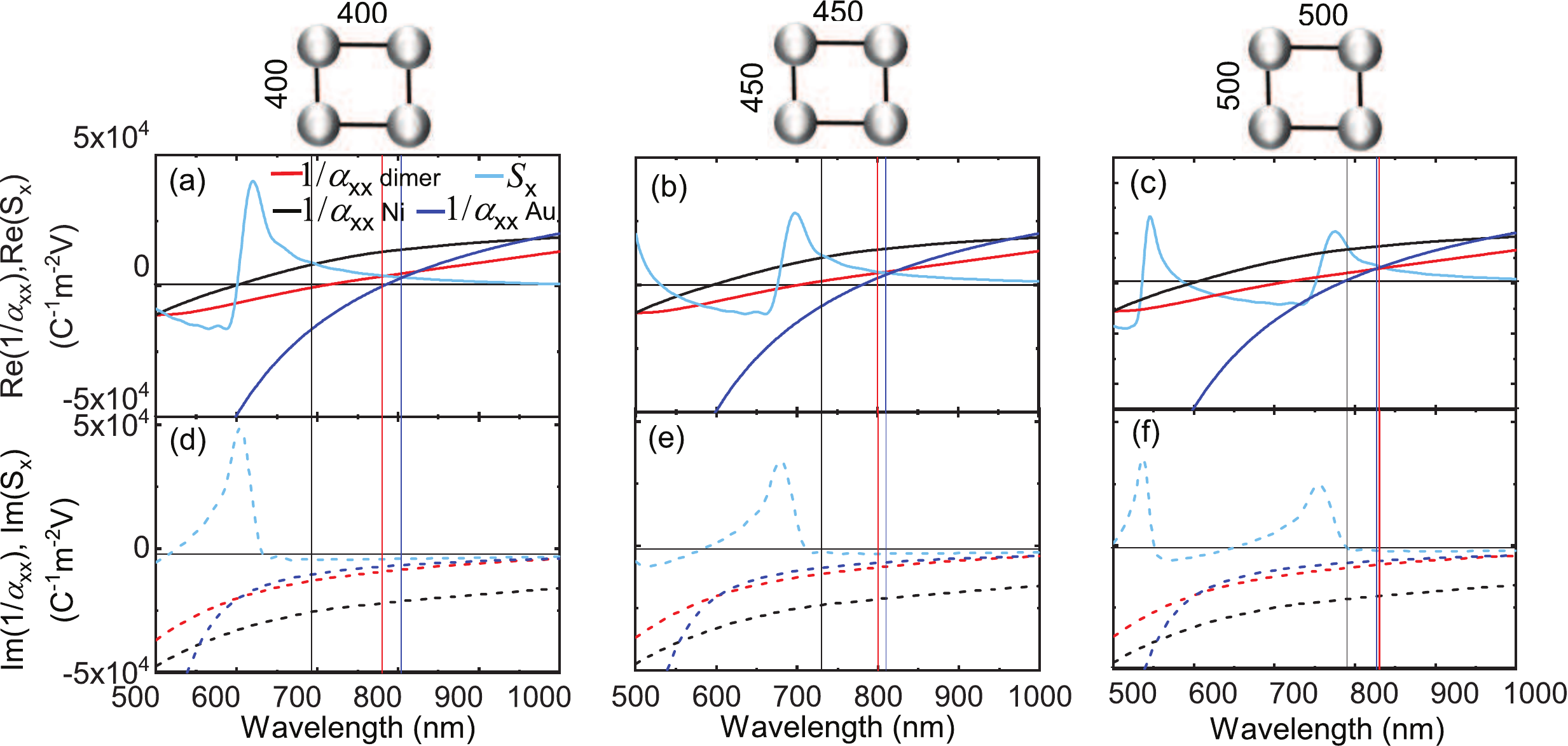}
	\caption{(a-c) Real and (d-f) imaginary parts of 1/$\alpha_\mathrm{xx}$ and $S_\mathrm{x}$. The 1/$\alpha_\mathrm{xx}$ curves depict the inverse polarizability of individual Ni/SiO$_2$/Au dimers and Ni and Au nanodisks. $S_\mathrm{x}$ solely depends on the lattice constant. Vertical lines indicate the wavelengths of SLR modes that combine Re(1/$\alpha_\mathrm{xx})-$ Re($S_\mathrm{x}$) = 0 and small Im($1/\alpha_\mathrm{xx})-$ Im($S_\mathrm{x}$). From these data, the effective polarizabilities of a periodic array can be calculated.}
	\label{fig:4}
\end{figure}

Figures \ref{fig:3}(a-c) show optical reflectance spectra for square arrays of dimers and Au and Ni nanodisks with lattice constants of 400 nm, 450 nm, and 500 nm. For these lattices, Rayleigh anomalies are observed at $\lambda_\mathrm{DO}=$ 610 nm, 680 nm, and 760 nm, respectively, in agreement with $\lambda_\mathrm{DO}=1.52a$. Because $\lambda_\mathrm{DO}$ only depends on the lattice constant, this feature is shared by all arrays. The signal minimum at the diffracted order is followed by a sharp increase of reflectance caused by the excitation of a collective SLR mode. Because the LSPRs of individual dimers and nanodisks are different, hybridization of these modes with the narrow DO produces SLRs with different lineshapes, resonance wavelengths, and intensities. For all particle types, the excitation of a SLR mode significantly enhances the reflectance in comparison to randomly distributed dimers and nanodisks (Fig. \ref{fig:2}(a)). The induced optical dipoles in lattices ($|p_\mathrm{eff,x}|$) are therefore stronger near the SLR wavelength. 
 
To analyze how excitations in the Au and Ni nanodisks of dimers contribute to the optical response of a periodic array, we consider the effective lattice polarizability along the incident electric field ($\alpha_\mathrm{eff,xx}$ in Eq. \ref{effective-polarizability-xx}). Parameter $\alpha_\mathrm{eff,xx}$ depends on the polarizability of individual dimers $\alpha_\mathrm{xx}$ and the lattice factor $S_\mathrm{x}$. In Fig. \ref{fig:4} we plot the real and imaginary parts of $1/\alpha_\mathrm{xx}$ and $S_\mathrm{x}$ for different lattice parameters. Data for the inverse polarizability of Ni and Au nanodisks are shown also. The effective polarizability of a nanoparticle lattice is resonantly enhanced when the real part of the denominator in Eq. \ref{effective-polarizability-xx}, 1/$\alpha_\mathrm{xx}-S_\mathrm{x}$, becomes zero. This condition corresponds to a crossing of the Re(1/$\alpha_\mathrm{xx}$) and Re($S_\mathrm{x}$) curves in Figs. \ref{fig:4}(a-c). The intensity and linewidth of the resulting SLR modes depend on the slope with which Re(1/$\alpha_\mathrm{xx}$) and Re($S_\mathrm{x}$) cross and the imaginary values of these parameters. For large Im($1/\alpha_\mathrm{xx})-$ Im($S_\mathrm{x}$) (Figs. \ref{fig:4}(d-f)), the SLRs are damped strongly. Since $S_\mathrm{x}$ solely depends on the lattice geometry, single particles only affect the excitation of SLRs through their inverse polarizability. Because Im(1/$\alpha_\mathrm{xx}$) can be written as $-$Im($\alpha_\mathrm{xx}$)/$|\alpha_\mathrm{xx}|^2$, it is approximated by $-1$/Im($\alpha_\mathrm{xx}$) close to the resonance condition (Re($\alpha_\mathrm{xx})\approx0$). For a dimer without gain $\alpha_\mathrm{xx}$ is positive and Im(1/$\alpha_\mathrm{xx}$) is negative. Consequently, the lattice factor $S_\mathrm{x}$ contributes to the damping of SLR modes if Im($S_\mathrm{x}$) is positive. In contrast, negative Im($S_\mathrm{x}$) counteracts the ohmic losses of individual nanoparticles, enabling the excitation of more narrow and intense SLRs. Because Im($S_\mathrm{x}$) changes sign from positive to negative at the DOs of a lattice, stronger SLR excitations are generated when the Re(1/$\alpha_\mathrm{xx}$) and Re($S_\mathrm{x}$) curves cross at $\lambda>\lambda_\mathrm{DO}$.  

The integration of Au into Ni/SiO$_2$/Au dimers, enlarges the polarizability of dimers in comparison to Ni nanodisks. Consequently, Im(1/$\alpha_\mathrm{xx}$) is smaller and SLR modes are less damped. Figures \ref{fig:4}(d-f) illustrate the large difference between Im(1/$\alpha_\mathrm{xx}$) of dimers and Ni nanodisks at relevant SLR wavelengths. To put some numbers on the resonant enhancement of the effective polarizability in our lattices, we compare the values of $|p_\mathrm{x}|$ in Fig. \ref{fig:2}(a) and $|p_\mathrm{eff,x}|$ in Figs. \ref{fig:3}(a,c). For single Ni/SiO$_2$/Au dimers and larger Ni nanodisks we measure $\alpha_\mathrm{xx}$(dimer)/$\alpha_\mathrm{xx}$(Ni disk) $\approx$ 1.4. In square lattices of the same particles $\alpha_\mathrm{eff,xx}$(dimer array)/$\alpha_\mathrm{eff,xx}$(Ni disk array) $\approx$ 3.2.       

In Fig. \ref{fig:4}(a) multiple crossings between Re(1/$\alpha_\mathrm{xx}$) and Re($S_\mathrm{x}$) are calculated for dimer and nanodisk arays with a lattice constant of 400 nm. However, only one of them, observed at $\lambda=690$ nm for Ni nanodisks, $\lambda=780$ nm for Au nanodisks, and $\lambda=805$ nm for Ni/SiO$_2$/Au dimers, coincides with a situation where Im($1/\alpha_\mathrm{xx})-$ Im($S_\mathrm{x}$) is small (Fig. \ref{fig:4}(d)). Consequently, one intense SLR mode is expected for these lattices, in agreement with the experimental spectra of Figs. \ref{fig:3}(a-c). Similar observations can be made for square arrays with lattice constants of 450 nm and 500 nm. The anticipated wavelengths of low-loss SLR modes for all particle types and lattice constants are indicated by vertical lines in Fig. \ref{fig:4}. Coupling between the diagonal (1,1) DO and LSPRs produces an additional SLR in lattices with $a=500$ nm. However, since Im(1/$\alpha_\mathrm{xx}$) is large at the wavelength of this mode, it appears much more damped in reflectance measurements. 

\begin{figure}[t]
	\centering
	\includegraphics[width=\linewidth]{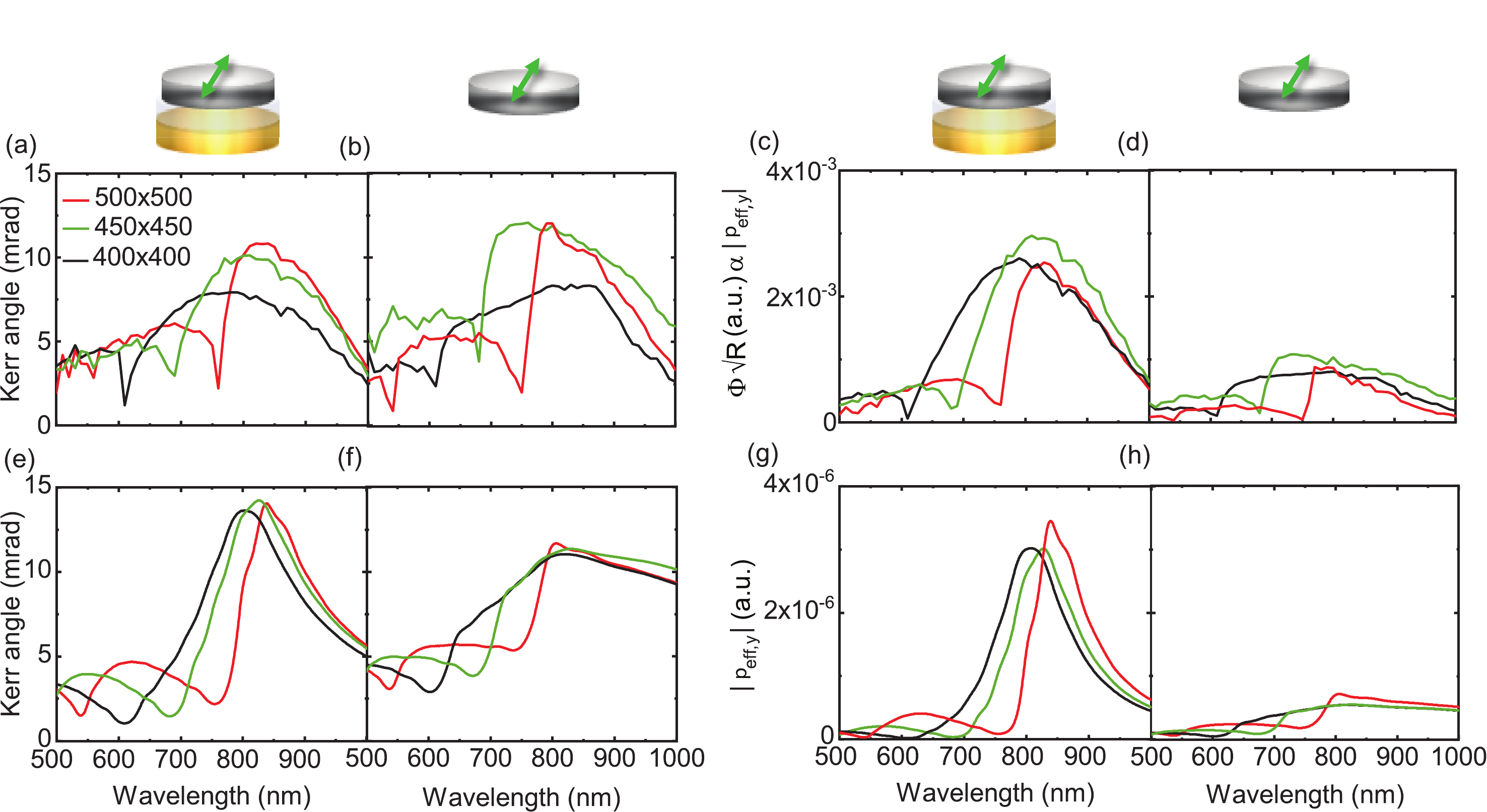}
	\caption{(a,b) Magneto-optical Kerr angle ($\Phi$) of square arrays of (a) Ni/SiO$_2$/Au dimers and (b) Ni nanodisks for three lattice constants. (c,d) Extracted values of $\Phi\sqrt{R}$ for the same lattices. This parameter, which is obtained from data in (a,b) and Figs. \ref{fig:3}(a,c), is proportional to the effective magneto-optical dipole ($|p_\mathrm{eff,y}|$). (e-h) Calculations of the magneto-optical Kerr angle ($|p_\mathrm{eff,y}/p_\mathrm{eff,x}|$) and $|p_\mathrm{eff,y}|$ for Ni/SiO$_2$/Au dimer and Ni nanodisk arrays.}
	\label{fig:5}
\end{figure}  

Another feature in the experimental reflectance spectra of Figs. \ref{fig:3}(a-c) that can be explained by considering Fig. \ref{fig:4} is the dependence of SLR wavelength on lattice constant. Because the slope of Re(1/$\alpha_\mathrm{xx}$) is particularly large for Au nanodisks, the crossing point between Re(1/$\alpha_\mathrm{xx}$) and Re($S_\mathrm{x}$) and, thus, the reflectance maximum only shifts slightly if Re($S_\mathrm{x}$) moves to higher wavelengths with increasing $a$. In contrast, smaller slopes of Re(1/$\alpha_\mathrm{xx}$) for dimers and Ni nanodisks result in stronger tuning of the SLR wavelength with lattice constant.      

To calculate the reflectance spectra of the different lattice ($|p_\mathrm{eff,x}|^2$), we insert data for 1/$\alpha_\mathrm{xx}$ and $S_\mathrm{x}$ from Fig. \ref{fig:4} into Eqs. \ref{effective-polarizability-xx} and \ref{lattice-dipole}. The results are shown in Figs. \ref{fig:3}(d-f). While our model calculations reproduce the main spectral features of the experimental curves, the resonances are more narrow. We attribute this discrepancy to inevitable imperfections in the experiments. For instance, we use a Gaussian beam with a finite wavelength range to excite our samples, while monochromatic plane waves are assumed in the calculations. Also, a finite distribution in the size and shape of the dimers and nanodisks (see Fig. \ref{fig:1}(a)) broadens the experimental resonances. 

\begin{figure}[t]
	\centering
	\includegraphics[width=0.7\linewidth]{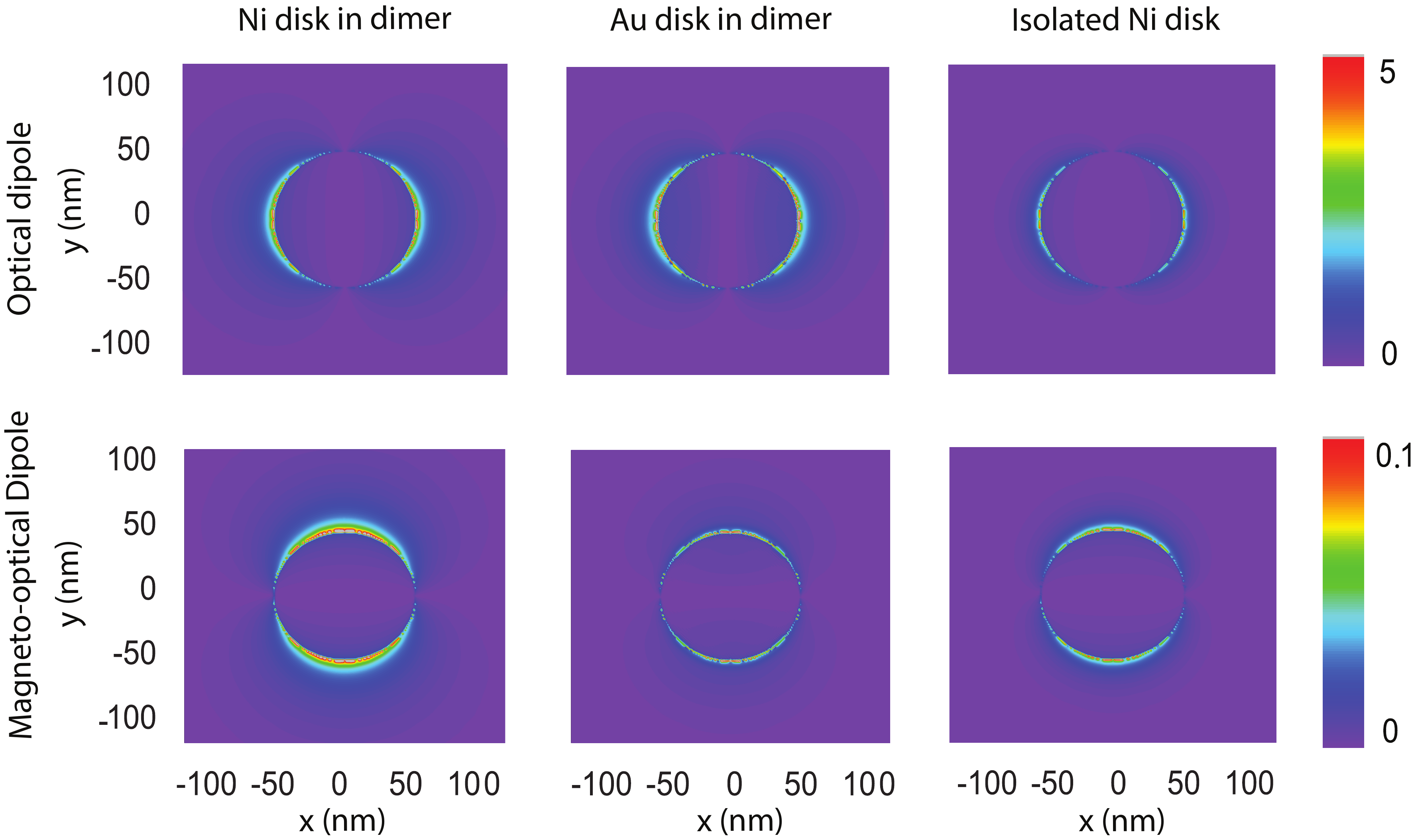}
	\caption{FDTD simulations of electric field distributions on top of the Ni and Au nanodisks of a dimer array and Ni nanodisks of a pure ferromagnetic lattice. The lattice constant is 400 nm. The disks are 15 nm thick and have a diameter of 110 nm. In the dimer array, the Ni and Au are separated by 15 nm SiO$_2$. Dipoles are excited at normal incidence with the electric field along the $x$-axis. The wavelength is set to $\lambda=780$ nm and the particles are embedded in an uniform medium with $n=1.5$.}
	\label{fig:6}
\end{figure}  

After establishing the optical response of different lattices, we now turn our attention to the magneto-optical activity of periodic Ni/SiO$_2$/Au dimer arrays. For comparison, we also discuss data for lattices with Ni nanodisks. Figure \ref{fig:5} shows the magneto-optical Kerr angle for square arrays with different lattice constants. Just like the optical reflectance measurements of Figs. \ref{fig:3}(a-c), the magneto-optical Kerr spectra are shaped by DOs (sharp minima) and SLR excitations (strong signal enhancements at $\lambda>\lambda_\mathrm{DO}$). The magnitude of the Kerr effect is comparable for periodic arrays of Ni/SiO$_2$/Au dimers and Ni nanodisks. According to Eq. \ref{Kerr-angle}, the off-diagonal to diagonal polarizability ratio ($|\alpha_\mathrm{eff,xy}/\alpha_\mathrm{eff,xx}|$) determines the Kerr angle of a lattice. Because the diagonal polarizability of the dimer array is much larger than that of the Ni lattice, we conclude that the off-diagonal polarizability of the dimer array must be similarly enlarged. To substantiate this claim, we multiply the Kerr data of Figs. \ref{fig:5}(a,b) with the square root of the reflectance spectra in Figs. \ref{fig:3}(a,c). The resulting parameter $\Phi\sqrt{R}$, shown in Fig. \ref{fig:5}(c,d), is proportional to the effective magneto-optical dipole ($|p_\mathrm{eff,y}|$) of the dimer and Ni lattices (Eq. \ref{MO-dipole}). Alike the effective optical dipole $|p_\mathrm{eff,x}|$ (Fig. \ref{fig:3}), the magneto-optical dipole $|p_\mathrm{eff,y}|$ of the Ni/SiO$_2$/Au dimer arrays is substantially stronger than that of pure Ni lattices. Thus, although the $|p_\mathrm{y}|$'s of individual dimers and larger Ni nanodisks are similar (Fig.  \ref{fig:2}(c)), the effective magneto-optical dipole is enhanced much more when dimers are ordered into periodic arrays. This result can be understood by considering Eq. \ref{effective-polarizability-xy} for the off-diagonal polarizabilities of a nanoparticle array. The effective off-diagonal polarizabilities of an array are directly proportional to the off-diagonal polarizabilities of the individual nanoparticles, which, as stated earlier, are similar for dimers and Ni nanodisks. However, the effective off-diagonal polarizability is resonantly enhanced when the real part of the denominator in Eq. \ref{effective-polarizability-xy} becomes zero. For square lattices with $\alpha_\mathrm{xx}=\alpha_\mathrm{yy}$ and $S_\mathrm{x}=S_\mathrm{y}$, this condition is met when the Re(1/$\alpha_\mathrm{xx}$) and Re($S_\mathrm{x}$) curves in Fig. \ref{fig:4} cross. Since resonances in $\alpha_\mathrm{eff,xx}$ and $\alpha_\mathrm{eff,xy}$ are determined by the same parameters in square arrays, the shapes of their optical and magneto-optical spectra are identical. Moreover, because Im(1/$\alpha_\mathrm{xx}$) is smaller for dimers than Ni nanodisks at the resonance wavelength, the magneto-optical Kerr angle is enhanced more by the excitation of an SLR mode in dimer arrays than in Ni lattices. Finally, we calculate the Kerr angle and magneto-optical dipole for both lattice types using the parameters of Fig. \ref{fig:4} and Eqs. \ref{effective-polarizability-xx}, \ref{effective-polarizability-xy}, \ref{lattice-dipole}, and \ref{Kerr-angle}. Results are plotted in Figs. \ref{fig:5}(e-h). The good agreements between the measured and calculated spectra demonstrate that our analytical model describes the physics of combined near- and far-field coupling in hybrid dimer lattices well.       

To visualize the excitation of SLRs in dimer and Ni nanodisk arrays, we performed finite-difference time-domain (FDTD) simulations. Results for square arrays with a lattice constant of 400 nm are shown in Fig. \ref{fig:6}. The data are obtained at $\lambda=780$ nm for both particle types. At this wavelength, the magneto-optical Kerr angle is enhanced by the excitation of a collective SLR mode (see Supplementary Note 3). Strong optical dipoles are directly excited by the incident electric field $E_\mathrm{i}$ along the $x$-axis. Through spin-orbit coupling in Ni nanodisks with perpendicular magnetization and near- and far-field interactions between Ni and Au disks, magneto-optical dipoles are induced along the $y$-axis in both Ni and Au. In agreement with our experiments and model calculations, the simulated dipole moments along $x$ and $y$ are larger in Ni/SiO$_2$/Au dimer arrays than in Ni nanodisk lattices. 

Finally, we consider the optical and magneto-optical response of rectangular dimer lattices with ${a_\mathrm{x}}\neq{a_\mathrm{y}}$. Based on our model, the optical reflectance of rectangular lattices depends on $\alpha_\mathrm{eff,xx}$. Because the lattice factor $S_\mathrm{x}$ peaks when $\lambda=1.52a_y$, the DO wavelengths are determined by the lattice constant along the $y$-axis. Consequently, only SLRs corresponding to this lattice period are expected in optical reflectance spectra. The same holds true for the magneto-optical response. While the denominator of $\alpha_\mathrm{eff,xy}$ (Eq. \ref{effective-polarizability-xy}) contains terms with $S_\mathrm{x}$ and $S_\mathrm{y}$, the magneto-optical Kerr angle is given by $|\alpha_\mathrm{eff,xy}/\alpha_\mathrm{eff,xx}|$ and thus

\begin{equation}
\Phi = \Bigg|\frac{\alpha_\mathrm{xy}}{\alpha_\mathrm{xx}\alpha_\mathrm{yy}(1/\alpha_\mathrm{xx} - S_\mathrm{y})}\Bigg|. 
\label{Kerr-angle-rectangle}
\end{equation}

\noindent Since $S_\mathrm{y}$ peaks when $\lambda=1.52a_x$, the SLR-enhanced magneto-optical response depends on the lattice parameter along the $x$-axis. This cross-dependence of the optical reflectance and magneto-optical Kerr angle on lattice constants $a_\mathrm{x}$ and $a_\mathrm{y}$, which has been observed previously for pure Ni lattices\cite{KAT-15}, is experimentally confirmed for dimers. The model prediction that the magneto-optical dipole $|p_\mathrm{eff,y}|$ of dimer lattices depends on both $S_\mathrm{x}$ and $S_\mathrm{y}$ is also verified by measurements. Experiments and model calculations on rectangular lattices are summarized in Supplementary Note 4.     

\section*{Conclusions}
We have experimentally and theoretically explored how plasmon resonances in hybrid Ni/SiO$_2$/Au dimer arrays compare to those of lattices that are made of Au or Ni nanodisks. Our results demonstrate that Ni/SiO$_2$/Au dimer arrays support more intense SLR modes than Ni lattices because the larger polarizability of individual dimer particles produces a stronger resonant enhancement of the effective lattice polarizability. The model that we present provides insight into the optical and magneto-optical response of ordered magnetoplasmonic dimers and offers clear directions on how to tailor the polarizability by material selection, variation of the particle size, or tuning of the lattice period or symmetry. 

\section*{Methods}
\subsection*{Sample preparation}
We fabricated the samples on glass substrates using electron-beam lithography. After spin-coating a polymethyl methacrylate (PMMA) layer and baking at 180$^\circ$C for 1 minute, the pattern was defined by exposing the resist layer to the electron beam. We developed the PMMA in a 1:3 methyl isobutyl ketone (MIBK):isopropanol (IPA) solution. Samples with pure Au or Ni nanodisks were fabricated by e-beam evaporation of a 15-nm-thick or 18-nm-thick film, followed by lift-off. For dimer samples, we first evaporated 1 nm Ti and 15 nm Au. After this, the samples were transferred to a magnetron sputtering system for the deposition of 15 nm SiO$_2$ (rf sputtering from a SiO$_2$ target). Finally, 15 nm of Ni was added and the stack was lift-off. We used SEM and atomic force microscopy to determine the nanodisk diameters. 

\subsection*{Optical and magneto-optical characterization}
\begin{figure}[t]
	\centering
	\includegraphics[width=0.7\linewidth]{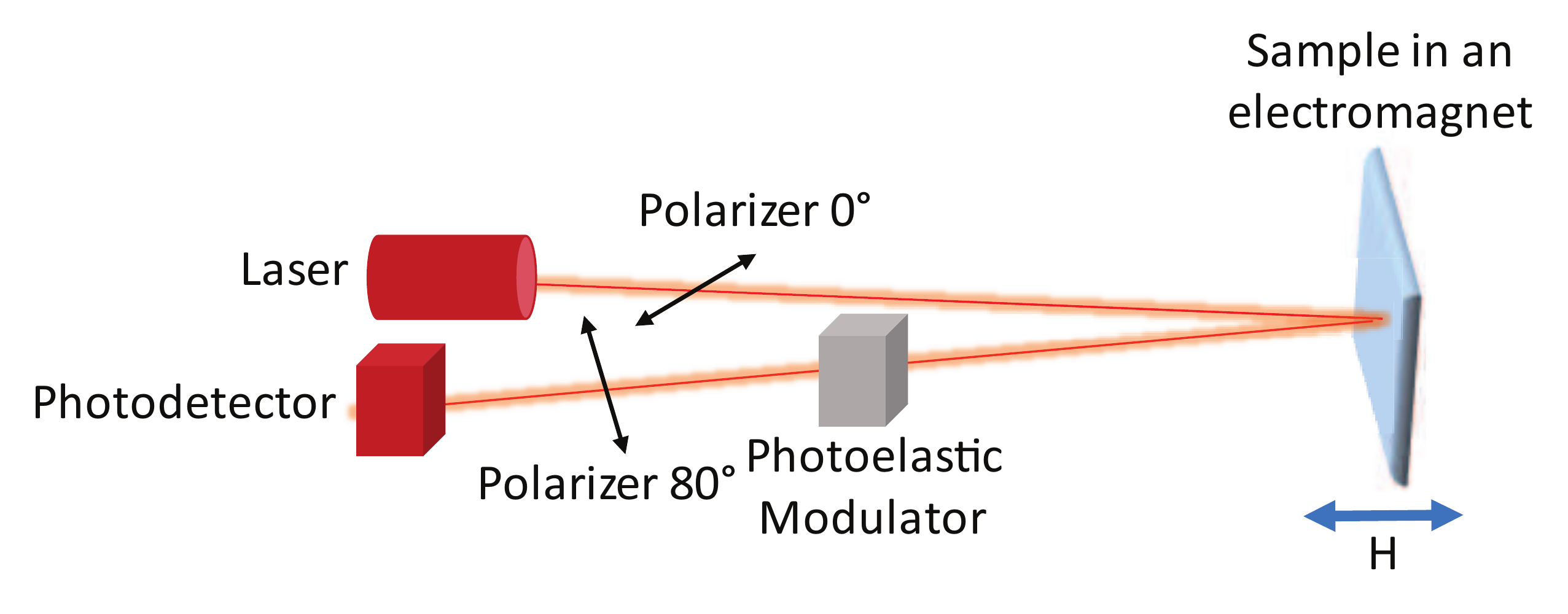}
	\caption{Schematic of the magneto-optical Kerr effect spectrometer. The setup consists of a broadband supercontinuum laser, polarizing and focusing optics, a photoelastic modulator, and a photodetector. We operate the instrument under normal incidence with linearly polarized light along the $x$-axis. A perpendicular magnetic field from an electromagnet saturates the magnetization of Ni nanodisks.}
	\label{fig:7}
\end{figure}

Optical reflectance and magneto-optical Kerr effect measurements were conducted with a Kerr spectrometer (Fig. \ref{fig:7}). The setup consisted of a broadband supercontinuum
laser (SuperK EXW-12 from NKT Photonics), polarizing and focusing optics, a photoelastic modulator (Hinds Instruments I/FS50), and a photodetector. The wavelength of the laser was tuned between 500 nm and 1000 nm. We used linear polarized light at normal incidence. During measurements, a $\pm$400 mT field from an electromagnet switched the magnetization of the Ni nanodisks between the two perpendicular directions. The Kerr rotation ($\theta$) and Kerr ellipticity ($\epsilon$) were simultaneously recorded by lock-in amplification of the modulated signal at 50 kHz and 100 kHz. From these data, we calculated the magneto-optical Kerr angle ($\Phi$) using 

\begin{equation}
\Phi = \sqrt{\theta^2+\epsilon^2}. 
\label{Kerr-angle-measurement}
\end{equation}

\subsection*{Finite-difference time-domain simulations}
Numerical simulations were carried out using finite-difference time-domain (FDTD) method. 400 nm $\times$ 400 nm unit cells comprising a vertical dimer made of 15-nm-thick Ni and Au nanodisks separated by 15 nm SiO$_2$ ($n=1.5$) or a single Ni disk of the same size were simulated. The disks diameters were set to 110 nm. Linearly polarized light was assumed to impinge along the sample normal from the Ni disk side. Periodic boundary conditions were applied at the edges of the simulation area. A uniform embedding medium with a dielectric constant of $n=1.5$ was used. Broadband reflectivity spectra were obtained by placing an electric field monitor 2 $\mu$m above the nanoparticles. Distributions of near-fields shown in Fig. \ref{fig:6} were calculated near the SLR wavelength. Magneto-optical effects were introduced in the FDTD simulations via off-diagonal terms in the permittivity tensor of Ni, while an isotropic dielectric function was assumed for Au. Distributions of magneto-optical dipolar fields were obtained by subtracting results for two perpendicular magnetization directions in Ni. In the simulations, these magnetic configurations were implemented by using opposite signs for the off-diagonal terms in the Ni permittivity tensor. 

\section*{Data Availability}

The data that support the findings of this study are available from the corresponding author upon request.


\section*{Acknowledgements}

This work was supported by the Academy of Finland (Grant No. 316857) and the Aalto Centre for Quantum Engineering. Lithography was performed at the Micronova Nanofabrication Centre, supported by Aalto University. P.V. acknowledges support from the Spanish Ministry of Economy, Industry and Competitiveness under the Maria de Maeztu Units of Excellence Programme – MDM-2016-0618.

\section*{Author Contributions}

S.P., M.K. and S.v.D. designed and initiated the research. S.P. fabricated the samples, conducted the measurements, and performed the model calculations with help from M.K. N.M. and P.V. conducted the FDTD simulations. S.v.D. supervised the project. All authors discussed the results. S.P., M.K., and S.v.D. wrote the manuscript, with input from N.M. and P.V.  

\section*{Additional information}

\textbf{Competing Interests:} The authors declare no competing interests. 

\end{document}